\documentclass[reprint,showpacs,preprintnumbers,superscriptaddress,multibbl,amsmath,amssymb,prl]{revtex4-1}
\usepackage[margin=0.5in]{geometry}                
\usepackage{geometry}
\usepackage{euscript,amsmath,amssymb,amsfonts,graphicx,bm}
\usepackage{hyperref}
\hypersetup{colorlinks=true,linkcolor=blue,citecolor=red}
\usepackage{epstopdf}
\usepackage{enumerate}
\usepackage{csquotes}
\usepackage{mathrsfs}
\usepackage{xcolor}
\usepackage{lineno}

\begin{document}

\title{A large deviation principle linking lineage statistics to fitness in microbial populations}

\date{\today}
\begin{abstract}
In exponentially proliferating populations of microbes, the population typically doubles at a rate less than the average doubling time of a single-cell  due to variability at the single-cell level. It is known that the distribution of generation times obtained from a single lineage is, in general, insufficient to determine a population's growth rate.  Is there an \emph{explicit} relationship between observables obtained from a single lineage and the population growth rate?  We show that a population's growth rate can be represented in terms of averages over isolated lineages. This lineage representation is related to a large deviation principle that is a generic feature of exponentially proliferating populations. Due to the large deviation structure of growing populations, the number of lineages needed to obtain an accurate estimate of the growth rate depends exponentially on the duration of the lineages, leading to a non-monotonic convergence of the estimate, which we verify in both synthetic and experimental data sets. 
\end{abstract}


\author{Ethan Levien}
\thanks{These two authors contributed equally}
\affiliation{School of Engineering and Applied Sciences, Harvard University, Cambridge, MA USA}
\author{Trevor GrandPre}
\thanks{These two authors contributed equally}
\affiliation{Department of Physics, University of California, Berkeley, CA USA}

\author{Ariel Amir}
\affiliation{School of Engineering and Applied Sciences, Harvard University, Cambridge, MA USA}

\maketitle


A key determinant of fitness in microbial populations is the population growth rate \cite{lin2017,lin2020,levien2019}.
 For organisms such as \emph{Escherichia coli} which undergo binary fission, the exponential growth rate of the population is determined by single-cell properties such as generation time, defined as the time from cell birth to division. In the simplest case where each cell in the population has a generation time of exactly $\tau_d$, the number of cells in the population, denoted $N(t)$, will grow as $N(t)\sim e^{\Lambda t}$, where the exponential growth rate, $\Lambda$ is given by $\Lambda = \ln(2)/\tau_d$. In reality, any clonal population of bacteria will exhibit a distribution of generation times due to a combination of factors, including intrinsic stochasticity of gene expression \cite{wang2011,duveau2018,elowitz2002,sughiyama2017}, asymmetric segregation of growth limiting resources at cell division \cite{chao2016,vedel2016,marantan2016,min2019} and environmental fluctuations \cite{claudi2014}. Together these factors will result in a distribution of generation times, $\psi(\tau_d)$, throughout the history of the population. The relationship between this distribution and the population growth rate, $\Lambda$, has been the subject of numerous studies. A key result is the \emph{Euler-Lotka equation} \cite{powell1956,lebowitz1974,lin2017,lin2020,levien2019,garciagarcia2019},
\begin{equation}\label{EL}
\frac{1}{2} = \int_0^{\infty}\psi(\tau)e^{-\Lambda \tau}d\tau, 
\end{equation}
which establishes a link between $\psi(\tau)$ and $\Lambda$. Equation \eqref{EL} is a generalization of a relation originally obtained by Euler and later rediscovered by Lotka \cite{Lotka1907}. 

Despite the elegant simplicity of the Euler-Lotka equation, it obscures the underlying relationship between the stochastic dynamics along single lineages and the population growth rate. The reason is that $\psi(\tau_d)$, like $\Lambda$, is a property of the population rather than an intrinsic property of individual cells and it is therefore unclear how differences in single cell dynamics are reflected in $\psi(\tau_d)$. Only in the special case where the generation time of a newborn cell is completely uncorrelated with its immediate ancestor, or \emph{mother}, does $\psi(\tau)$ correspond to the distribution of generation times along a single-lineage \cite{lebowitz1974,lin2017}. In this limit one can, in principle, infer $\Lambda$ without access to the entire population.  In contrast, when generation times are correlated between mother and daughter cells, the distribution of generation times, $f(\tau)$, along a single lineage no longer contains enough information to deduce the growth exponent $\Lambda$ using equation (\ref{EL}).  Such correlations emerge naturally through feedback mechanisms and are required to maintain homeostasis of cell sizes  \cite{amir2014,ho2018}: if intrinsic stochasticity within a cell causes it to grow abnormally large, the daughter cell will tend to have a shorter generation time in order to account for the additional size inherited from its mother. This leads to negative correlations between mother and daughter cells. There are numerous other mechanisms which can generate correlations between mother and daughter cells. For example, environmental fluctuations can induce positive correlations between mother and daughter cells \cite{levien2019,powell1956}.

\begin{figure}[h!]
 \includegraphics[scale=0.3]{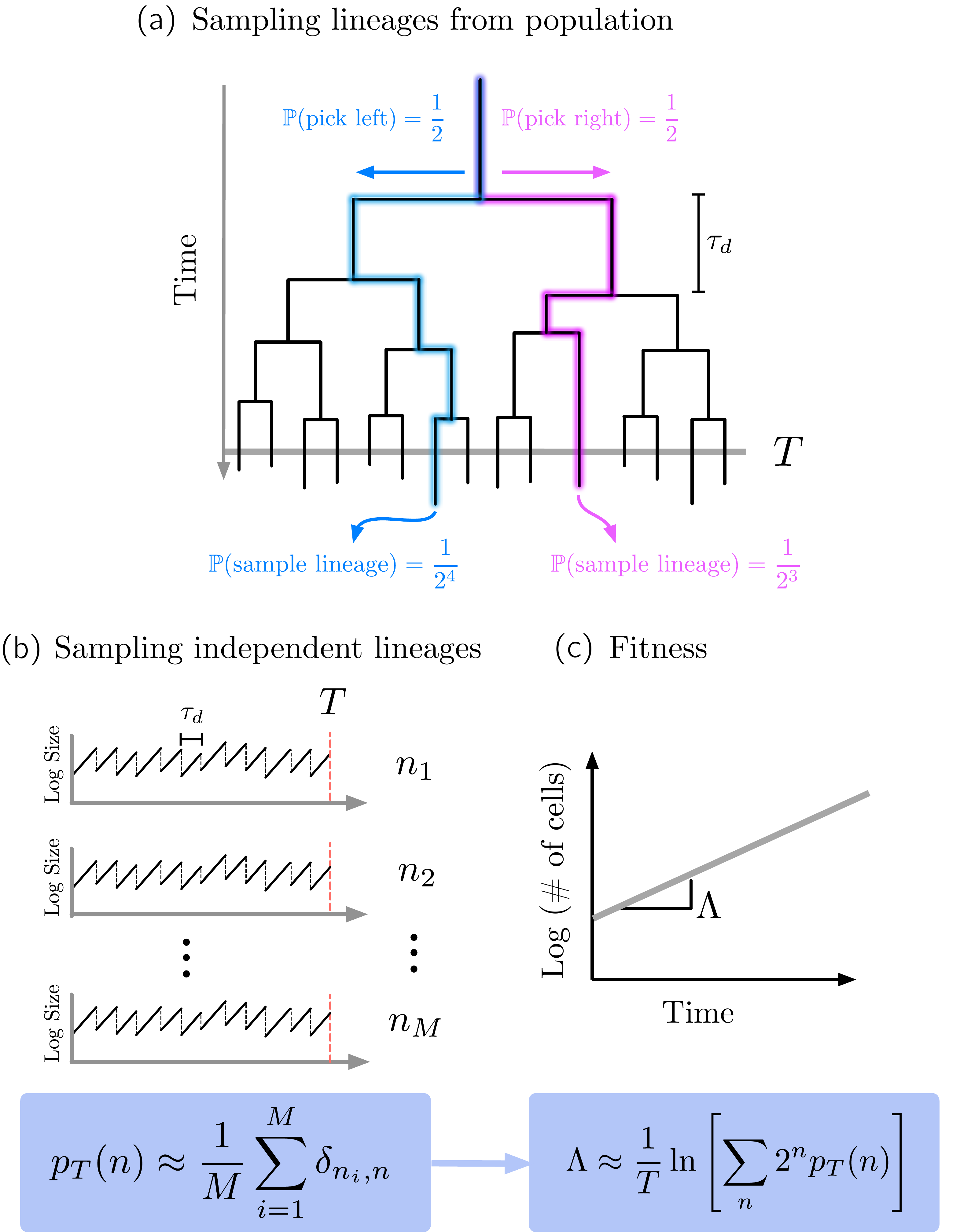}
\caption{(a) A population tree starting from a single ancestor. The distinction is made between single lineages (highlighted) and a population (black). Lineages can be sampled by traveling down the tree and randomly selecting a daughter cell at the end of each branch. The probability of selecting any specific lineage with $n$ divisions is $2^{-n}$.   (b) $M$ independent lineages of length $T$. For the $i$th lineage, $n_i$ is the number of cell divisions along that lineage. For each lineage we have shown the cell size, which typically increases exponentially between divisions, as a function of time. The lineage division distribution can be approximated from these independent lineages by recording the division events and using the highlighted formula.  (c)  A growing population of cells from which one can compute the fitness directly by counting the number of cells as a function of time (or utilizing equation \eqref{EL}). Using the lineage distribution of divisions we can obtain the fitness from independent lineages. }\label{fig:1}
\end{figure}

The discrepancy between the statistics of a lineage and those from the entire population, which determine the population's fitness, raises the question of how to quantify fitness from data obtained from a single lineage, or a collection of independent lineages; see Figure \ref{fig:1}. Such data is typically obtained 
from mother machine experiments \cite{robert2010}. In these experiments, independent lineages are tracked for long periods of times in highly controlled conditions. Mother machine experiments enable detailed measurements of single-cell dynamics that would be impossible in bulk conditions. In contrast, bulk experiments can be used to probe population-level dynamics and measure fitness, but are blind to the physiological details at the microscopic level \cite{robert2010}. 
Here, we present a lineage representation of the population growth rate that connects the population dynamics to the statistics along a single lineage, or a collection of independent lineages. Our lineage representation reveals that a large deviation principle underlies population growth. This generalizes existing variational principles for the population growth rate (e.g. the results of \cite{wakamoto2012}) to the context where generation times are correlated.

{\emph{Lineage representation}}.
A lineage-based representation of the population growth rate that is independent of the model specifics can be derived using the \emph{division distribution}, denoted $p_T(n)$, which can be obtained from an exponentially growing population as follows. Suppose a population of cells is grown for a time $T$ and assume that we have access to the generation times of individual cells and the genealogical relationships between cells, as shown in Figure \ref{fig:1}. We emphasize that our only assumption is that the population grows exponentially as $T \to \infty$. We can randomly sample a lineage from the tree by starting from the ancestral cell in the population and randomly selecting one of its daughter cells with equal probability to obtain the next cell in the lineage. Repeating this procedure yields a single lineage, as shown by the highlighted paths in Figure \ref{fig:1}.

If $N(T)$ is the number of cells in the population at time $T$, then there are exactly $N(T)$ lineages, as each cell in the final population corresponds to a distinct lineage. However, by randomly selecting a lineage in the \emph{forward} manner described above, lineages with more divisions are less likely to be selected, since each division decreases the chance that we will travel down that specific path through the tree. 
In particular, the probability of drawing any specific lineage from the tree is $2^{-n}$. It follows that the empirical division distribution, denoted $\hat{p}_T(n)$, of observing exactly $n$ divisions in a lineage sampled using this procedure is given by \cite{wakamoto2012,nozoe2017}
\begin{equation}
\hat{p}_T(n) = 2^{-n}N(n,T),
\end{equation}
where $N(n,T)$ is a random variable representing the number of lineages with $n$ divisions in a specific realization of a growing population. Note that $\hat{p}_T(n)$, is also a random variable, and will therefore differ between different realizations of the population tree. By averaging over many realization of the tree, we obtain the division distribution:
\begin{equation}
p_T(n) \equiv \langle \hat{p}_T(n) \rangle_{\rm trees}.
\end{equation}
It is important to remember that $p_T(n)$ is distinct from what has been called the \emph{retrospective distribution}, defined as the probability of observing $n$ divisions in a lineage obtained by uniformly sampling a cell from the population at time $T$ and following its ancestors back in time \cite{nozoe2017}. 

Given a specific realization of the population, the total number of cells can be represented in terms of the empirical division distribution as
\begin{equation}
 N(T)  = \sum_{n} N(n,T) = \sum_{n}2^{n}\hat{p}_T(n).
\end{equation}
Averaging over many realization of the population now yields the average population size in terms of the division distribution
\begin{equation} 
\langle N(T) \rangle_{\rm trees}   = \sum_{n}2^{n}p_T(n).
\end{equation}
Importantly, $p_T(n)$ is not a random variable that depends on a specific realization of the tree, rather it is an intrinsic property of lineages and has no information about the correlations between sister cells. It can therefore be obtained by running many \emph{independent} lineages: if we have a collection of $M$ independent lineages of length $T$ and observe $n_i$ divisions in the $i$th lineage, then 
\begin{equation}
p_T(n) = \lim_{M \rightarrow \infty} \frac{1}{M}\sum_{i}^{M} \delta_{n_i,n}; 
\end{equation}
see Figure \ref{fig:1} (b). 
The long-time population growth rate, defined as 
\begin{equation}\label{Lambda_def}
\Lambda \equiv   \lim_{T \to \infty}\frac{1}{T}\ln N(T),
\end{equation}
can now be related to an average over independent lineages according to the \emph{lineage representation},
\begin{equation}\label{Lambda_lin}
\Lambda  = \lim_{T \to \infty}\frac{1}{T}\ln \left \langle  2^n\right \rangle_p. 
\end{equation}
Here, the angular brackets denote an average over $p_T(n)$. The definition of $\Lambda$ in equation \eqref{Lambda_def} is justified in SM section 1, where we have shown that this limit is self-averaging (i.e.  in the long-time limit, only a single realization of the population is needed to obtain $\Lambda$). The lineage representation in equation (\ref{Lambda_lin}) establishes a relationship between the lineage dynamics and the population fitness. A similar formulation was used in \cite{nozoe2017} to quantify how selection acts on an observable in a growing population; however, our formulation differs in that the average is taken over \emph{independent} lineages rather than lineages from a single growing population. 

In order to apply the lineage representation of the population growth rate to real data, we must develop an understanding of how quickly it converges in the number of lineages, $M$, and the duration of each lineage, $T$. As we will show below, some care needs to be taken when selecting $M$ and $T$, as the lineage representation has a non-monotonic convergence in $T$. However, before presenting our convergence analysis, we establish a relationship between the lineage representation and the large deviation principle underlying the growth process. This structure is best introduced with the simple example below. 


{\emph{Explicit calculation of $p_T(n)$ for discrete Langevin model}}.
We now perform an explicit calculation of $p_T(n)$ for a specific model in which generation times undergo a discrete Langevin process along a lineage, this model is referred to as the \emph{random generation time} model in the literature \cite{lin2017,lin2020}.  In this model, the generation time $\tau$ of a cell is related to its mother's generation time, $\tau'$, according to 
\begin{equation}
\tau = \tau_0 (1-c) + \tau' c+ \xi
\end{equation}
where $\xi$ is a Gaussian with mean zero and variance $\sigma_{\xi}^2$. The parameter $c$ controls the strength of correlations between mother and daughter cells, and for $c=0$ we retrieve the classical Bellman-Harris branching process \cite{harris1952}. It can be seen that the average generation time along a lineage is $\langle \tau \rangle = \tau_0$. The population growth rate for this model has previously been obtained using a recursive approach in ref \cite{lin2020}. Here, we are able to derive the same result by implementing the lineage representation. Additionally, we obtain the complete distribution of divisions which elucidates the underlying large deviation structure of the population growth process. 

We proceed by writing the probability that the $n$th division will occur along a lineage at  time $T$:
\begin{align}\label{qn0}
\begin{split}
q_T(n) &= \int_0^{\infty} \int_{0}^{\infty} \cdots \int_0^{\infty} \delta\left(\sum_i t_i - T \right)\\
&\quad \times \prod_i \frac{1}{\sqrt{2\pi \sigma_{\xi}^2}} e^{-(t_i-ct_{i-1}-\tau_0(1-c))^2/2\sigma_{\xi}^2}dt_i. 
\end{split}
\end{align}
Note that this is distinct from $p_T(n)$, the probability of observing exactly $n$ divisions \emph{before} reaching time $T$. However, since we are interested primarily  in the large deviations, the two distributions are interchangeable.  By making a change of variables and performing a Gaussian integral (see SM section 2), equation \eqref{qn0} leads to the simple formula
\begin{equation}\label{pn-rgt}
p_T(n) =  
Ke^{-n\frac{(1-c)^2}{2\sigma_{\tau}^2}\left(\tau_0-\frac{T}{n} \right)^2},
\end{equation}
where $K$ is a normalization constant independent of $n$ and $\sigma_{\tau}^2 = \sigma_{\xi}^2/(1-c^2)$ is the variance in $\tau$ taken over a single lineage. Notice that the distribution of divisions is not Gaussian but due to the quadratic dependence of the exponent on $T/n$, that of the inverse of the number of divisions is approximately Gaussian. We elaborate on this fact in SM section 4. 

The exponential form of equation (\ref{pn-rgt}) along with equation (\ref{Lambda_lin}) suggests that the population growth rate is dominated by a particular value of $n$ which maximizes the exponent of $2^np_T(n)$. Treating $n$ as a continuous variable and solving for $n$ in $\frac{\partial}{\partial n}\left[n \ln2 + \ln p_T(n)\right] = 0$ yields the dominant number of divisions
\begin{equation}\label{critical_n}
n_c = T\Big /\sqrt{\langle \tau \rangle^2 - \frac{2\ln(2)\sigma_{\tau}^2}{(1-c)^2}}.
\end{equation}  
Note that in the limit where $\sigma_{\tau}^2 \to 0$, we find $n_c = T/\langle  \tau \rangle$, which is the number of divisions corresponding to the average generation time. 
Substituting only the dominant value of $n$ from equation \eqref{critical_n} into equation (\ref{Lambda_lin}) gives us the formula for the bulk population growth rate:  $\Lambda = n_c\ln(2)/T + \ln p(n_c,T)/T$. After some simplification, we obtain
\begin{equation}\label{Lambda_example}
\Lambda=  \frac{2 \ln(2)/\langle \tau \rangle}{1 + \sqrt{1-2 \ln(2)\frac{\sigma_{\tau}^2}{\langle \tau \rangle^2}\frac{1+c}{1-c}}},
\end{equation}
which is in agreement with previous computations using an alternative approach \cite{lin2020}. From equation (\ref{Lambda_example}), we can see how the three model parameters, namely $\langle \tau\rangle$, $\sigma_{\tau}^2$ and $c$, affect population growth. In particular, growth is increased when $\sigma_{\tau}^2$ and $c$ are increased, while increasing $\langle \tau\rangle$ decreases growth. 

{\emph{Large deviation principle}}. In order to connect the observation of the previous section to large deviation theory, we introduce the \emph{time averaged division rate} $\gamma = n/T$, so that the distribution of division rates given by equation \eqref{pn-rgt} can be expressed as
\begin{equation}\label{ldp}
p_T(\gamma) \propto e^{-TI(\gamma)}
\end{equation}
with 
\begin{equation}\label{rate_f}
    I(\gamma)= \frac{\gamma (1-c)^2}{2\sigma_{\tau}^2}\left(\tau_0-1/\gamma\right)^2.
\end{equation}
The exponential dependence of $p_T(\gamma)$ on $T$ is known as a \emph{large deviation principle} and suggests that for large $T$ averages over $p_T(\gamma)$ are dominated by a single value of $\gamma$ \cite{touchette2009large}. For the remainder of this paper we will assume this large deviation principle is satisfied.

 In SM section 3, we show that when $I''(\langle \gamma \rangle_p) \gg 1$ we can make a Gaussian approximation of $p_T(\gamma)$ to obtain 
\begin{equation}\label{cumu_exp2}
\Lambda \approx \frac{\ln(2)}{\langle \tau \rangle} + \frac{T\ln(2)^2\sigma_{\gamma}^2}{2}
\end{equation}
with $\sigma_{\gamma}^2 = 1/TI''(\langle \gamma \rangle_p)$. 
This illustrates a central result of the large deviation formulation; namely, regardless of the model specifics, corrections to the population growth rate due to variation in generation times scale inversely with the curvature of the large deviation rate function.

\begin{figure}[h!]
\includegraphics[scale=0.55]{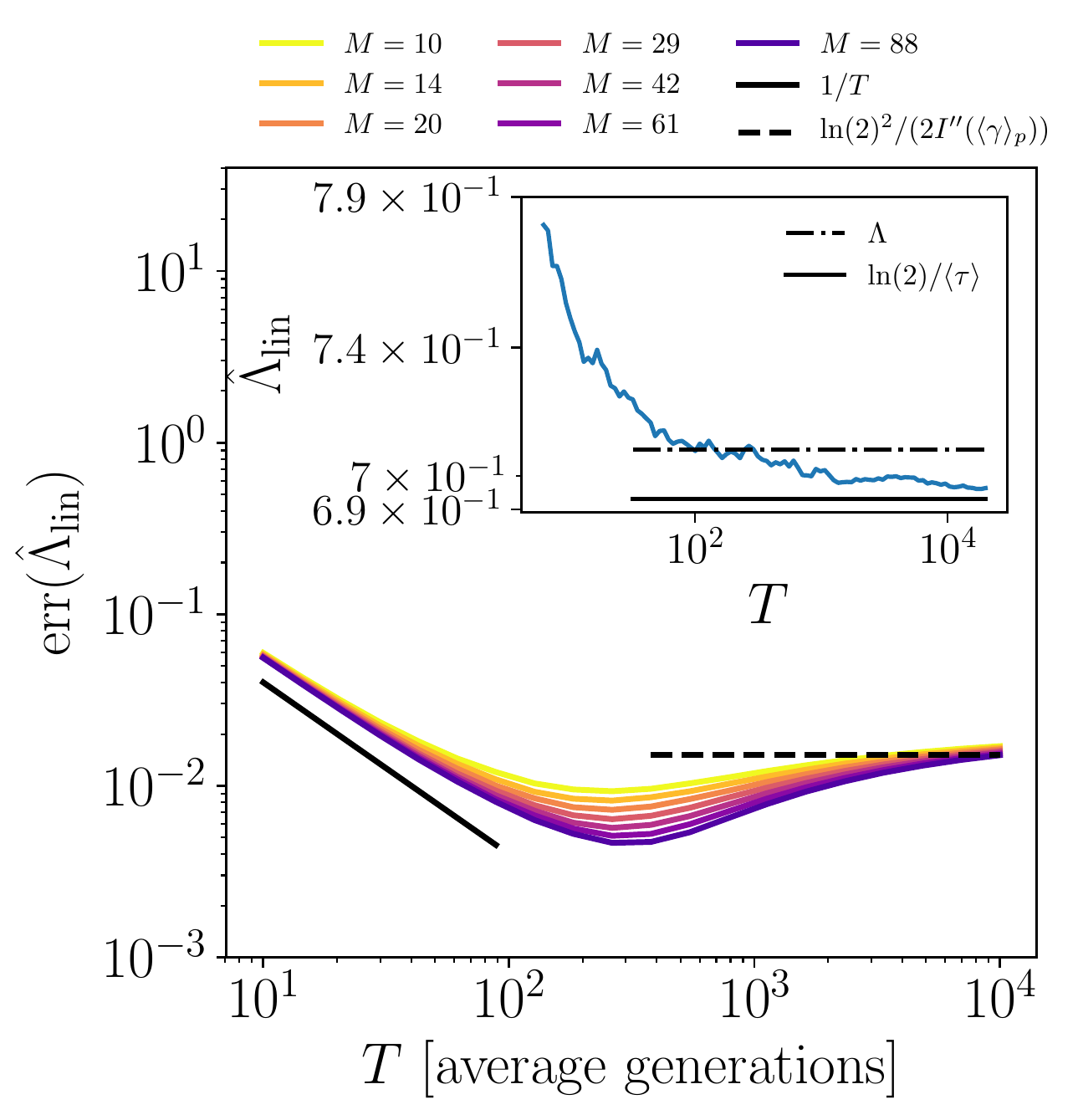}
\caption{Convergence of the error from the lineage representation as a function of the lineage durations, $T$, for different numbers of lineages, $M$. Data was generated from lineage simulations of the Langevin model with $\langle \tau \rangle  = 1$, $c=0.2$ and $\sigma_{\tau} = 0.2$.  Here it can clearly be seen that the error initially scales as $1/T$, eventually increasing to approach the limit imposed on the sampling error by the large deviation rate function; see equation \eqref{uc}. The inset shows the lineage representation $\hat{\Lambda}_{\rm lin}$ as a function of $T$ using $M = 80$ lineages. This plot is noisy because only a single ensemble of lineages is used, in contrast ${\rm err}$ in the main plot is computed by averaging over many ensembles of lineages. 
}\label{fig:2}
\end{figure}

{\emph{Convergence of lineage representation.}}
We now address the question: How accurately can we estimate $\Lambda$ given $M$ lineages with durations $T$? To quantify the accuracy of an estimate, denoted $\hat{\Lambda}_{\rm lin}$, we use the averaged squared deviation 
\begin{equation}
{\rm err}(\hat{\Lambda}_{\rm lin}) ^2= \left \langle  \left( (\hat{\Lambda}/\Lambda -1 \right)^2  \right\rangle_{\mathcal E}.  
\end{equation}
Here, the average $\langle \cdot \rangle_{\mathcal E}$ represents the average over many realization of the ensemble of $M$ lineages of duration $T$, not to be confused with the averages elsewhere that are taken over lineages within an ensemble. Two distinct factors contribute to the error: First, the estimate of $\Lambda$ obtained from the lineage representation will be subject to a systematic error resulting from the fact that given an infinite number of lineages each with a finite duration $T$, the lineage representation approximates the \emph{arithmetic mean} fitness at time $T$: $\Lambda_{T,a} = 1/T \ln \langle N \rangle$. This is distinct from $\Lambda$. (In fact, it is not even the correct measure of fitness for a population grown over a finite period of time, which is given by the \emph{geometric mean} $1/T  \langle  \ln N \rangle$; see ref. \cite{lewontin1969} for an explanation of why this is.)  We refer to this error as \emph{finite duration error}, and as we have shown in the SM section 4, it will scale inversely with $T$. 

The second factor contributing to ${\rm err}(\hat{\Lambda})$ is sampling error in the approximation of the average $\langle2^n \rangle_p$ from a finite number of lineages.
As shown in the SM section 4, when
 \begin{equation}\label{samp_scale}
\frac{1}{\langle 2^{n} \rangle_p }  \sqrt{\frac{{\rm var}(2^n)}{M}}= \sqrt{\frac{2^{T\ln(2)/I''(\langle  \gamma \rangle_p)} - 1}{M}}\ll 1
 \end{equation}  the contribution of the sampling error to ${\rm err}(\hat{\Lambda}_{\rm lin})$ will grow exponentially in $T$ for any fixed $M$, eventually dominating the error resulting from finite lineage durations. As $T$ becomes large, the distribution of $\gamma$ becomes much more narrow, so an ever-increasing number of lineages are needed to sample the variation in $\gamma$; however, as we have seen in equation \eqref{cumu_exp2}, knowledge of the variation is needed to resolve the effects of generation time variability on population growth. In the long-time limit, all information about the variation is lost for finite $M$ and the lineage representation simply retrieves the zeroth order term in equation \eqref{cumu_exp2}:
 \begin{equation}\label{uc}
 \lim_{T \to \infty}\hat{\Lambda}_{\rm lin} =  \ln(2)\langle  \gamma \rangle_p.
 \end{equation}
 This demonstrates that the $T$ and $M$ limits do not commute, because as we have already seen, if we first take $M\to \infty$ the lineage representation converges to the exact population growth rate. As a result, there is a ``goldilocks effect": if $T$ is too small the estimate will be inaccurate due to the finite duration error, while if $T$ is too large we encounter the limit given by equation \eqref{uc}. The best estimate is in fact obtained by using an intermediate $T$ where both effects are minimized.   This prediction is validated numerically for the Langevin model in Fig. \ref{fig:2}. We have also generated the same data for a more biophysically realistic model of cell growth (the cell-size regulation mode \cite{amir2014}), and found the results are qualitatively similar; see SM section 5.

How much data do we need to be confident we are not encountering the limit given by equation \eqref{uc}? 
The sampling error will have a negligible effect on the estimate when equation \eqref{samp_scale} is satisfied. This condition can be rewritten as 
\begin{equation}\label{Mbest}
M\gg 2^{T\ln(2)/I''(\langle  \gamma \rangle_p)}. 
\end{equation}
 This implies that the number of lineages needed to avoid encountering the sampling error grows exponentially with the duration of the lineages and the generation time variance. In order to be confident that the finite duration error is small enough to resolve the second term in equation \eqref{cumu_exp2}, we must select $T \gg I''(\gamma_c)$. This means that $T\ln(2)/I''(\langle  \gamma \rangle_p)$ is a large quantity. For example, if we want to ensure that the finite duration error is an order of magnitude smaller than the generation time variance, a safe value of $M$ will generally be larger than $2^{10 \times \ln(2) } \approx 120$.   Existing data sets obtained from mother machines contain on the order of one hundred lineages, making the application of the lineage representation plausible. In SM section 6, we have explored the applications of the lineage algorithm to mother machine data, where we have found that the dependence of $\hat{\Lambda}_{\rm lin}$ on $T$ is qualitatively consistent with the theory presented above and Fig. \ref{fig:2}.

{\emph{Discussion}}.
Experimental advances over the last few decades have made it possible to observe the stochastic dynamics of growth and division in bacteria with increasing levels of precision \cite{robert2010,luro2019,camsund2019}. These observations have revealed universal principles underlying microbial growth, such as the adder mechanisms for maintaining homeostasis of cell sizes \cite{amir2014,ho2018,cadart2018,eun2017,logsdon2017}. Bulk experiments in which bacteria are grown exponentially or in competition assays can be used to compare the fitness of different strains, and in principle elucidate how these physiological differences map to fitness. However, the equivalence between growth in bulk experiments and those used to observe single-cell traits remains unclear, because of the different environments which cells are subjected to in these experiments.
In this paper, we have presented a lineage representation that links single-lineages to the population growth.

We have also found that a large deviation principle underlies the population dynamics which applies to the distribution of division rates among lineages in a growing population. This implies that the population becomes dominated by lineages with a certain optimal division rate.  This idea generalizes an \emph{optimal lineage} principle introduced by Wakamoto et al. which was used to calculate the population growth rate within the context of a model with uncorrelated generation times \cite{wakamoto2012}.  Using the large deviation framework, we have quantified exactly how much data is needed to resolve the effects of cell-to-cell variability on population growth from single lineages.  
We expect that this work will serve as a guide for future experimental studies seeking to link single-cell observations to fitness.

\vspace{0.3cm}
 \begin{acknowledgments}
We thank Jie Lin and Somenath Bakshi for helpful discussions related to this work. We thank Jane Kondev for providing feedback on the manuscript. We acknowledge funding support from National Science Foundation under Award DMS-1902895 (EL), Californian Alliance Research Exchange with NSF Grants 1647273/1742065/1306595/1306683/1306747/1306760 (TG) and NSF CAREER 1752024 (AA). 
\end{acknowledgments}

\bibliography{lineage_alg.bib}
\onecolumngrid
\newpage

\begin{appendix}
\begin{center}
\textbf{\large SUPPLEMENTAL MATERIALS}
\end{center}

\renewcommand{\theequation}{S\arabic{equation}}
\setcounter{equation}{0}
\setcounter{figure}{0}
\setcounter{table}{0}
\setcounter{page}{1}
\makeatletter
\renewcommand{\theequation}{S\arabic{equation}}
\renewcommand{\thefigure}{S\arabic{figure}}

\section{1. Argument that $\Lambda$ is self-averaging}\label{app:Lambda_def}
We show that, provided fluctuations in generation times have a finite correlation time, 
\begin{equation}
\Lambda_T = \frac{1}{T}\ln N(T)
\end{equation}
converges to a deterministic value in the large $T$ limit. This will establish that equation \eqref{Lambda_def} is an appropriate definition for the long-term fitness. We begin by defining averages of $\Lambda_T$ in two ways: First, we consider the geometric mean
\begin{equation}
\Lambda_{T,g} = \frac{1}{T}\langle \ln N(T) \rangle
\end{equation}
where the average is taken over many realizations of the population. Second, we consider the arithmetic mean fitness:
\begin{equation}
\Lambda_{T,a} = \frac{1}{T} \ln \langle N(T) \rangle. 
\end{equation}
We will say that $\Lambda_T$ is self-averaging if the geometric and mean fitness converge to the same value in the long-time limit. 
The distinction between mean and geometric fitness has been shown to be important when considering populations growing in the presence  of environmental stochasticity. In this context, it is often the case that the two are not equal and that the geometric fitness is the more appropriate measure of the long-term viability of a population; see ref. \cite{lewontin1969} for a detailed discussion of this point. In the context of branching processes, the fact that $\Lambda_T$ is self-averaging has been established for the Bellman-Harris branching process model; see ref. \cite{harris1952}. For completeness, we provide an argument that is slightly more general, and allows for mother and daughter cells to have correlated generation times. 

Setting $N(T) =  \langle N(T) \rangle  + dN$, we have 
\begin{align}
\begin{split}
\Lambda_{T,g} &= \frac{1}{T}\left\langle \ln\left[ \langle N(T) \rangle \left( 1 + \frac{dN}{ \langle N(T) \rangle } \right) \right] \right\rangle  \\
& \approx \Lambda_{T,a} - \frac{1}{2T} \frac{\langle dN^2 \rangle }{\langle N(T)^2 \rangle }\\
&=  \Lambda_{T,a} - \frac{1}{2T}{\rm CV}^2_{N}
\end{split}
\end{align}
where ${\rm CV}^2_{N}$ is the coefficient of variation of $N(T)$.  We now argue that ${\rm CV}^2_{N}$ converges to a constant as $T$ becomes large. To do this, we write down the differential equation for the probability distribution of $N(t)$, denoted $P(N,t)$.  Let $\alpha(t)$ be the per unit time, per capita probability of a cell dividing. The time dependence in $\alpha(t)$ comes from the fact that the distribution of ages will take some time to converge to its steady-state \cite{Jafarpour2019,Jafarpour2018}.  Since the division rates of individual cells are age-dependent, the per capita division rates throughout the population will depend on time.   It has previously been shown that if there is any variability in generation times and the fluctuations in generation times have a finite correlation time, the distribution of ages will eventually become time-invariant \cite{lebowitz1974,levien2019}. 
Hence $\alpha(t)$ will converge to a constant, $\bar{\alpha}$, in the long-time limit. 
Thus, after a long period of time, the dynamics of $P(N,t)$ are approximated by
\begin{equation}\label{Pmaster}
\frac{d}{dt}P(N,t) \approx \bar{\alpha} (N-1) P(N-1,t)  - \bar{\alpha} N P(N,t).
\end{equation}

To proceed we analyze the moments of $N$ taking the initial condition to be $\langle N(t_0)^k \rangle=n_k$. It is important that we are only considering the dynamics after an initial transient, otherwise equation \eqref{Pmaster} will not be a valid approximation. For the first moment of $N$, we find
\begin{align}
\begin{split}
\frac{d}{dt}\langle N \rangle 
&=\bar{\alpha}\sum \left[(N+1) N P(N,t) -N^2 P(N,t) \right]=\bar{\alpha} \langle N \rangle
\end{split}
\end{align}
implying $\langle N \rangle =n_1 e^{(t-t_0) \bar{\alpha}}$ and $\bar{\alpha} = \Lambda_{T,a}$.
Similarly, 
\begin{align}
\begin{split}
\frac{d}{dt}\langle N^2 \rangle &\approx \bar{\alpha}  \sum N^2 \left[(N-1) P(N-1,t)  -N P(N,t)\right]\\
&=\bar{\alpha} \left[ n_1 e^{t\bar{\alpha} } + 2 \langle N^2\rangle \right]. 
\end{split}
\end{align}
Solving this ODE with $\langle N^2 \rangle$ gives 
\begin{align}
\langle N^2 \rangle  &\approx e^{(t-t_0)\bar{\alpha}} \left[n_2 e^{(t-t_0) \bar{\alpha}} + n_1\left(e^{(t-t_0) \bar{\alpha}}-1 \right) \right]
\end{align}
Since both $\langle N^2 \rangle$ and $\langle N \rangle ^2$ grow asymptotically as $e^{2\bar{\alpha}t}$, the coefficient of variation of $N(t)$ will converge to a constant. In particular, 
\begin{equation}
{\rm CV}_N^2(t) = \frac{n_2}{n_1^2} + \frac{1-e^{-\bar{\alpha}(t-t_0)}}{n_1} \to \frac{n_2 + n_1}{n_1^2}.
\end{equation}

\begin{figure}[h!]
\includegraphics[scale=0.6]{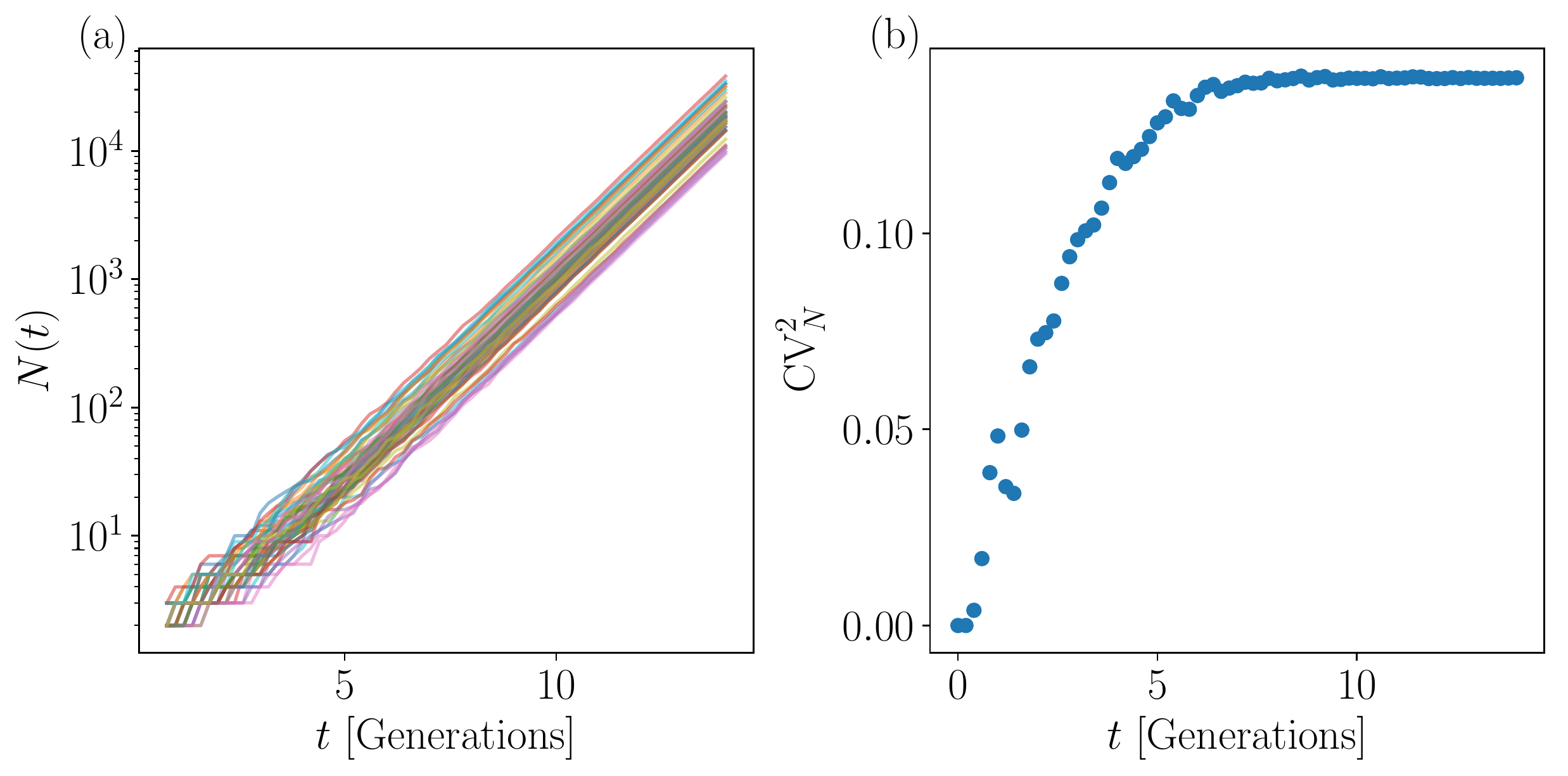}
\caption{ (a) Many realizations of $N(t)$ on a log plot. For all realizations, the slope of $\ln N(t)$ is eventually constant and independent of the realization. Thus $\Lambda$ is well-defined. (b) The coefficient of variation of $N(t)$ as a function of time. The CV is computed from $500$ simulations of $N(t)$.   Data was generated from simulations of the full population for the Langevin  model with $\langle \tau \rangle  = 1$, $c=0.2$ and $\sigma_{\tau} = 0.2$.  
}\label{fig:SI1}
\end{figure}

 It follows that 
\begin{equation}
\lim_{T \to \infty} \Lambda_{T,a} = \lim_{T \to \infty} \Lambda_{T,g}, 
\end{equation}
which implies 
\begin{equation}
\Lambda = \lim_{T \to \infty}\Lambda_T
\end{equation}
is non-random. The predictions of this section are verified in Figure \ref{fig:SI1}, where we have shown that the exponential growth rate of $N(t)$ is deterministic and ${\rm CV}_N$ converges to a constant, although our calculation does not give an explicit formula for this constant. 

\section{2. Details in derivation of $p_T(n)$ for Langevin model}

Here we provide details on the calculation of $p_T(n)$ for the Langevin model. 
Starting with equation \eqref{qn0} of the main text, we replace the $\delta$ function with $\frac{1}{2\pi} \int_{-\infty}^{\infty} e^{i\omega x}d\omega$ and introduce the constant $b = \tau_0(1-c)$. This leads to an equivalent form
\begin{align}
\begin{split}
&q_T(n) \propto \int_{-\infty}^{\infty} e^{-i\omega T}\left[ \int_0^{\infty} \cdots  \int_0^{\infty} \prod_{j=0}^n e^{i\omega t_j}e^{(t_j-ct_{j-1}-b)^2/2\sigma_{\xi}^2}dt_j \right]d\omega.
\end{split}
\end{align}
 We can complete the square in the integrand and define the constants $\bar{b} = b + i\omega 2 \sigma_{\xi}^2/(2(1-c))$ and $d = i\omega c2\sigma_{\xi}^22/(1-c)$. This leads to the integral
\begin{align}
q_T(n) \propto \int_{-\infty}^{\infty} e^{-i\omega T} \left[ \int_0^{\infty}  \cdots  \int_0^{\infty} \prod_{j=0}^{n} e^{(t_j-ct_{j-1}-\bar{b})^2/2\sigma_{\xi}^2} e^{(\bar{b}^2 - b^2)n /2 \sigma_{\xi}^2} e^{t_n d}dt_j   \right] d\omega.
\end{align}
In order to evaluate this integral, we make the transformation $x_i = t_i - ct_{i-1}-\bar{b}$. Note that the Jacobian from this change of variables is 1. When the noise in generation times is small enough for the contribution from the negative part of the integral to be negligible, we can make the approximation that integrals are evaluated over the real line. This leads to
\begin{align}\label{qn}
\begin{split}
&q_T(n) \propto \int_{-\infty}^{\infty} e^{-i \omega T}e^{-(\bar{b}^2 - b^2)n/2\sigma_{\xi}^2} \left[ \int_{-\infty}^{\infty} \cdots \int_{-\infty}^{\infty}\prod_{j=0}^ne^{-x_j^2/2\sigma_{\xi}^2}e^{d t_n}dx_j \right]d\omega.
\end{split}
\end{align}
Also, notice that $t_n$ can be expressed as a sum over all the $x_i$s as
\begin{equation}
t_n = x_n + cx_{n-1}+c^2x_{n+2} + \cdots. 
\end{equation}
For $|c|<1$, which is always the case in our model, this only has an affect on a finite number of terms, thus the $n$th term can be neglected for large $n$. We find that 
\begin{equation}
q_T(n) =  \int_{-\infty}^{\infty}e^{-i\omega T}e^{n(\bar{b}^2 - b^2)/2\sigma_{\xi}^2}d\omega. 
\end{equation}
Using the identity 
\begin{equation}
\frac{\bar{b}^2-b^2}{2\sigma_{\xi}^2} = \frac{-\omega^2 2\sigma_{\xi}^2}{4(1-c)^2} + \frac{i \omega b}{1-c}
\end{equation}
and computing the integral, we see that the formula for $q$ is given by equation \eqref{pn-rgt} of the main text. To obtain $p_T(n)$ from $q_T(n)$, we note that  $q_T(n)$ is the probability of observing $n$ divisions weighted by the frequency of passing through age zero, which is time invariant  (see \cite{lebowitz1974,levien2019}), hence in the long time limit the large deviation rate functions for these two distributions will be equivalent.

\section{3. Variational formulation of population growth rate}
Using the large deviation formulation, we can obtain an alternative representation of $\Lambda$. First, we can take a Laplace Transform of equation \eqref{ldp} of the main text, yielding 

\begin{equation}\label{laplace}
    \left\langle e^{T\gamma \beta}\right\rangle_{p}=\int e^{T(\gamma \beta-I(\gamma))} d{\gamma}.
\end{equation}
Setting $\beta=\ln 2$, we see that the population growth rate is given by a saddle point over $\gamma$ written as
\begin{equation}\label{Lambda_laplace}
    \Lambda=\max_{\gamma}\left[\gamma \ln 2-I(\gamma) \right].
\end{equation}
This formulation is equivalent to equation \eqref{Lambda_lin} of the main text, but makes an explicit connection between the growth rate and the Large deviation rate function.  
In addition, we can cast equation \eqref{Lambda_laplace} as a cumulant expansion of equation \eqref{laplace} with $\beta=\ln(2)$. This gives
\begin{equation}\label{cumu_exp}
    \Lambda=\left\langle\gamma\right\rangle_p\ln(2)+\frac{T\ln(2)^2}{2}\sigma_{\gamma}^2+\cdot \cdot \cdot,
\end{equation}
where $\sigma_{\gamma}^2 = 1/TI''(\langle \gamma\rangle_p)$ is the variance of $\gamma$ with respect to $p_T(\gamma)$. 
We see that the first term in equation \eqref{cumu_exp}  is simply the growth rate calculated from the doubling time of a single lineage, $\ln(2)/\langle\tau\rangle$, while including the second order term gives us equation \eqref{cumu_exp} becomes equation \eqref{cumu_exp2} of the main text.  In the context of the Langevin model,  the curvature of the large deviation rate function is given by $I''(\langle \gamma \rangle_p) \approx \left(1-c\right)^2 \langle\tau \rangle ^3/\sigma_{\tau} ^2$. It is straightforward to check that equation \eqref{cumu_exp} is consistent with equation \eqref{Lambda_example} to leading order in $\sigma_{\tau}^2/\langle  \tau \rangle^2$ and $c$. If either one of these parameters is large, then the distribution $p_T(\gamma)$ will be too broad for the first term in the cumulant expansion to give a good approximation.

\section{4. Analysis of convergence}
We analyze the convergence of the lineage representation by first considering the limit in which we have an infinite number of lineages ($M \to \infty$), each with a large, but finite duration $T$. In this case, the lineage representation gives the arithmetic mean fitness, $\Lambda_{T,a}$. This can be obtained from the saddle point approximation,
\begin{align}
\left \langle 2^{\gamma T} \right \rangle & = \frac{K}{T}\int_0^{\infty}e^{-TI(\gamma)+ \gamma T\ln(2)}d\gamma\\
&\approx  \frac{K}{T}\sqrt{\frac{2 \pi}{T I''(\gamma_c)}}e^{-TI(\gamma_c) + \gamma_cT\ln(2)}. 
\end{align}
Note that the factor $1/T$ outside the integral comes from the Jacobian when we change variables from $n$ to $\gamma$. It follows that 
\begin{align}\label{LT1}
\Lambda_{T} &= \frac{1}{T}\ln \left \langle 2^{\gamma T} \right \rangle  = \Lambda + \frac{1}{T}\ln \left[\frac{K}{T^{3/2}}\sqrt{\frac{2 \pi}{ I''(\gamma_c)}} \right]. 
\end{align}
In order for the normalization of $p_T(n)$ to hold, $K$ must grow as $\sqrt{T}$.
We therefore conclude that the convergence in $T$ is (ignoring logarithmic terms) $O(1/T)$. The specific value of the prefactor will depend on the initial transient dynamics when the population is small, and therefore cannot be computed using the large deviation estimates, which are valid when  $T$ is large.

\begin{figure}[h!]
\includegraphics[scale=0.6]{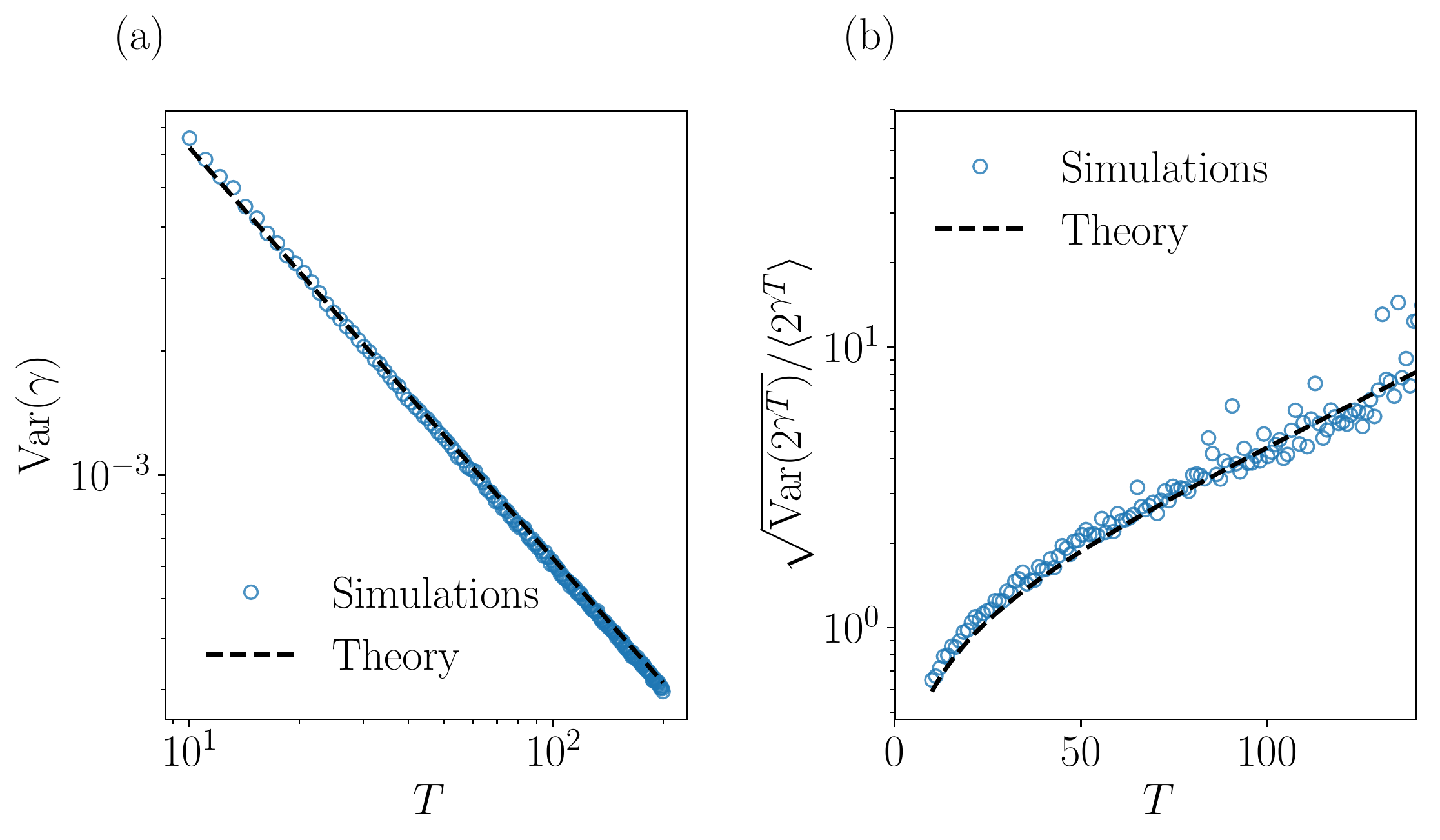}
\caption{Validation of the predictions for the statistics of $\gamma$ using the Langevin model.  (a) A demonstration of the linear decay in the variance of $\gamma$ with time as predicted by equation \eqref{sigma_gamma}. (b) The coefficient of variation of $2^{\gamma T}$ compared to the theoretical prediction of equation \eqref{var_2gamT} obtained from a Gaussian approximation.   Data was generated from lineage simulations of the Langevin  model with $\langle \tau \rangle  = 1$, $c=0.2$ and $\sigma_{\tau} = 0.2$.
}\label{fig:SI2}
\end{figure}

Now consider the estimate obtained from a finite number of lineages. For large $M$, by applying the central limit theorem, we get 
\begin{equation}\label{2gam_clt}
\left\langle 2^{\gamma T} \right\rangle_M  \approx \left\langle 2^{\gamma T} \right\rangle + \sqrt{\frac{{\rm Var}(2^{\gamma T})}{M}}\eta
\end{equation}
where $\eta$ is a standard normal variable. 
The rate function can be approximated as a Gaussian by writing 
\begin{equation}
    I(\gamma)\approx \frac{1}{2} I''(\langle  \gamma \rangle_p)(\gamma-\langle  \gamma \rangle_p)^2=\frac{(\gamma-\langle  \gamma \rangle_p)^2}{2T\sigma^2_{\gamma}}. 
\end{equation} 
Recall that $\langle \gamma \rangle_p$ is the average of $\gamma$ with respect to $p_T(\gamma)$.  In this approximation, $\gamma$ is a normally distributed random variable with variance 
\begin{equation}\label{sigma_gamma}
\sigma_{\gamma}^2 = \frac{1}{TI''(\langle  \gamma \rangle_p)}.
\end{equation}
  For a random variable $X$ with mean $\mu$ and variance $\sigma^2$, $\langle e^X \rangle  = e^{\mu + \sigma^2/2}$. Approximating $\gamma$ with a Gaussian and applying this identity to $2^{\gamma T}$, we have
\begin{align}
\langle 2^{\gamma T}\rangle 
&\approx  e^{T\ln (2)\langle \gamma \rangle  + T^2\ln(2)^2\sigma_{\gamma}^2/2} ,\\
\left\langle \left(2^{\gamma T}\right)^2\right\rangle &= e^{2T\ln (2)\langle \gamma \rangle  + T^2\ln(2)^22\sigma_{\gamma}^2}\\
&\approx \langle 2^{\gamma T}\rangle^2 2^{ T^2\ln(2)\sigma_{\gamma}^2} = \langle 2^{\gamma T}\rangle^2 2^{ T\ln(2)/I''(\langle  \gamma \rangle_p)}.
\end{align}
Hence for the variance, we have
\begin{align}\label{var_2gamT}
{\rm Var}(2^{\gamma T}) &=\left\langle \left(2^{\gamma T}\right)^2\right\rangle - \langle 2^{\gamma T}\rangle^2\\
&=     \left\langle 2^{\gamma T} \right\rangle^2\left(2^{T\ln(2)/I''(\langle  \gamma \rangle_p)} - 1 \right). 
\end{align}
The predictions for the statistics of $\gamma$ are validated numerically in Figure \ref{fig:SI2}.

When $\sqrt{{\rm Var}(2^{\gamma T})/M}/\langle2^{\gamma T}\rangle \ll 1$, we have
\begin{align}
\hat{\Lambda}_{\rm lin} &= \frac{1}{T}\ln \left(  \left\langle 2^{\gamma T} \right\rangle +  \sqrt{\frac{{\rm Var}(2^{\gamma T})}{M}} \eta\right)\\
&=  \frac{1}{T}\ln  \left\langle 2^{\gamma T} \right\rangle + \frac{1}{T}  \ln \left( 1+ \frac{1}{ \left\langle 2^{\gamma T} \right\rangle}\sqrt{\frac{{\rm Var}(2^{\gamma T})}{M}}\eta \right)\\
&\approx \Lambda_T +\frac{1}{T\left\langle 2^{\gamma T} \right\rangle}\sqrt{\frac{{\rm Var}(2^{\gamma T})}{M}} \eta \\
&\approx \Lambda + \underbrace{\frac{1}{T}\ln \left[\frac{K}{T^{3/2}}\sqrt{\frac{2 \pi}{ I''(\gamma_c)}} \right]}_{\text{finite duration error}}
 + \underbrace{\frac{1}{T}\sqrt{\frac{2^{T\ln(2)/I''(\langle  \gamma \rangle_p)} - 1}{M}}\eta}_{\text{sampling error}}.\label{Llin_shorttime}
\end{align}
From this calculation, we see that when equation \eqref{Mbest} of the main text is satisfied  the sampling error is $O(M^{-1/2})$. However, for any fixed $M$ the the second term in equation \eqref{Llin_shorttime} grows exponentially in $T$. Hence for large enough $T$ the expansion breaks down and we need to obtain the error in a different way. To this end, we consider the limit of a finite number of lineages, each with an infinite duration. Let $\gamma_i$ be the empirical division rate along the $i$th of $M$ lineages. In the limit $T \to \infty$, 
\begin{equation}
\lim_{T\to \infty}\gamma_i = \langle \gamma \rangle_p.   
\end{equation}
Therefore, if we first take the $T$ limit, rather than having a distribution of $\gamma$ each lineages gives the same value of $\gamma_i = \langle \gamma \rangle$. As a result, the lineage algorithm simply gives 
\begin{equation}
\Lambda_{\rm lin} = \ln(2)\langle  \gamma \rangle_p. 
\end{equation}
We conclude that for any finite $M$ the lineage representation is not effective for large $T$. In particular, the representation only gives reasonable results if equation  \eqref{Mbest}  is satisfied. This also establishes that the $T$ and $M$ limits of the lineage representation do not commute, and it is only if we first take the $M$ limit that the estimate converges. 

\section{5. Test of convergence on cell-size regulation model}
Here, we test the lineage representation on a more biophysically realistic model of microbial growth. This model has been used previously in the literature to understand how cells maintain homeostasis of their sizes \cite{ho2018,amir2014}. The central assumption of the cell-size regulation model is that cells grow exponentially at the single-cell level and divide when they reach a size $v_{\rm div}$ depending on their size at birth, $v_{\rm birth}$. Since cells grow exponentially, the generation time of a cell satisfies $v_{\rm div} = v_{\rm birth}e^{\lambda \tau}$, or 
\begin{equation}\label{csr_tau}
\tau = \frac{1}{\lambda}\ln \left(\frac{v_{\rm div}}{v_{\rm birth}} \right).
\end{equation}  
Here, $\lambda$ is the single-cell growth rate. For simplicity, we will assume that cells divide symmetrically; thus $v_{\rm birth}$ is obtained by dividing the cell's mother's size at division by $2$.  Phenotypic variability is introduced in the form of both variability in single-cell growth rates and noise in the division volumes. To implement this, we take the growth rate $\lambda$ and division volume $v_{\rm div}$ of a cell to obey \cite{lin2020} 
\begin{align}
\begin{split}
	\ln \lambda &= \ln \langle \lambda \rangle + \eta_{\lambda}\\
	v_{\rm div} &= 2(1-\alpha)v_{\rm birth} + 2\alpha v_0 + \eta_{v}
\end{split}
\end{align}
where $\eta_{\lambda}$ and $\eta_v$ are independent normally distributed random variables with variances $\sigma_{\lambda}^2$ and $\sigma_{v}^2$, respectively. The first equation captures the fact that growth rates approximately follow a log-normal distribution for small noise, while the second equation tells us how cells decide when to divide based on their volume. The parameter $\alpha$ determines the cell-size regulation strategy. When $\alpha=1$  (known as a ``sizer" strategy), cells divide at a critical size, while when $\alpha =1/2$ (known as an ``adder" strategy) cells add a constant size $v_0$ between birth and division.  We refer to refs \cite{ho2018,lin2017} for an in-depth discussion of the cell-size control model and its implications for population growth.

\begin{figure}[h!]
\includegraphics[scale=0.7]{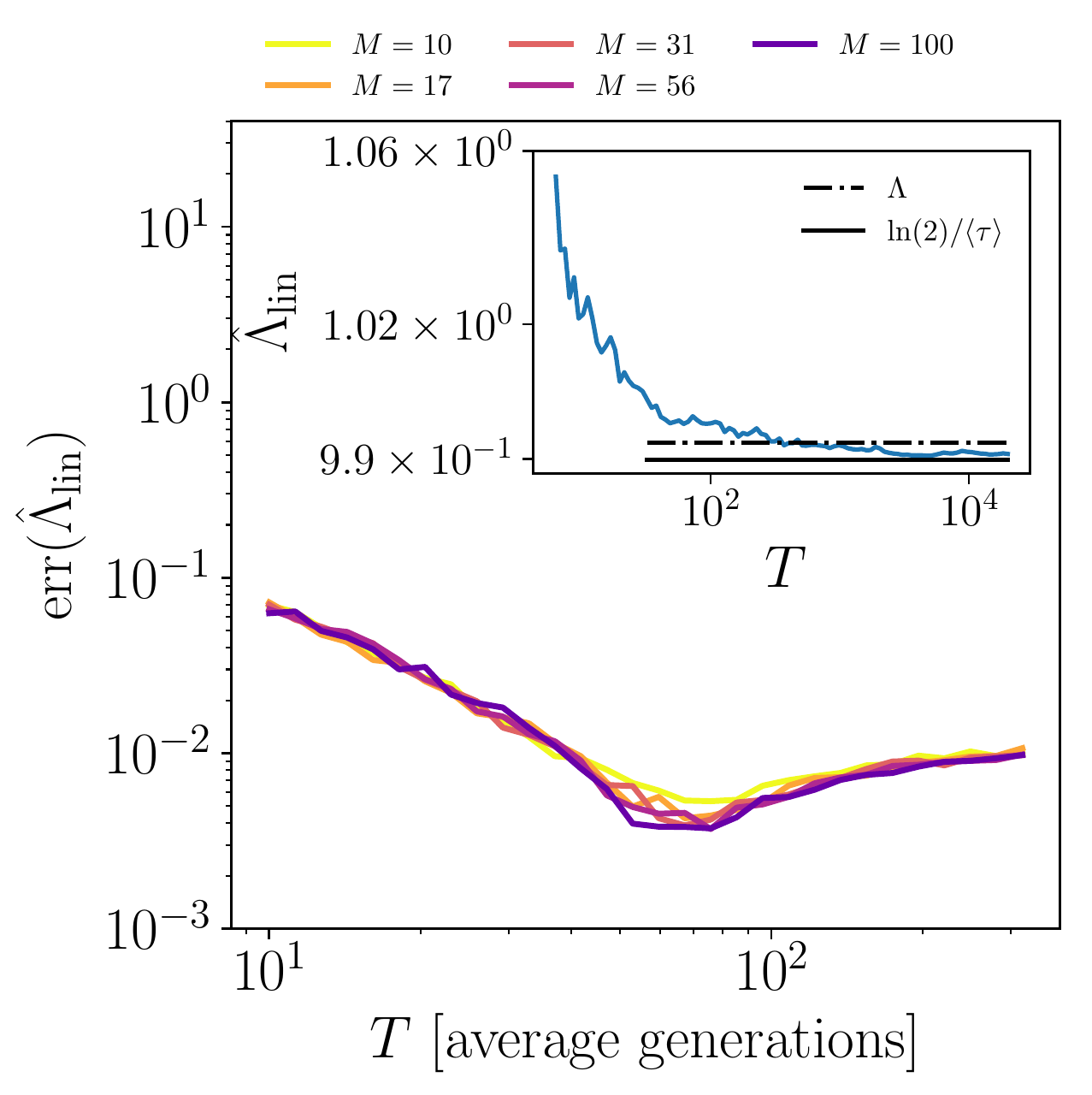}
\caption{Here we have run simulations in the same manner as Fig. \ref{fig:2} of the main text, but used the cell-size regulation model instead of the Langevin model to compute the generation times. Parameter values used are $\sigma_v = 0.2$, $\sigma_{\lambda} = 0.1$ and $\alpha=1/2$. 
}\label{fig:SI3}
\end{figure}

The convergence of the lineage representation for the cell-size regulation model is shown in Figure \ref{fig:SI3}, which should be compared to Figure \ref{fig:2}. We see that the behavior of the error in $T$ and $M$ is qualitatively similar to the the Langevin model, with the convergence being non-monotonic in $T$. 

\section{6. Application to experimental data}
Here, we apply the lineage representation to the mother machine data from ref \cite{tanouchi2017}. 
In these experiments, \emph{E. coli} were grown in three different temperatures. Each experiment resulted in a collection of roughly $M \approx 80$ independent lineages (recordings of division times) with durations on the order of $100$ generations. 
In order to explore how the duration of the lineages affects our estimate, we have computed $\hat{\Lambda}_{\rm lin}$ for different values of $T$ by truncating the lineages. Along with the lineage algorithm, we have computed two other measurements of growth: First, we have computed the first order term in equation \eqref{cumu_exp2}, $\ln(2)/\langle  \tau\rangle$. Second, we have computed the average single-cell elongation rate along the lineages. If $v_{\rm birth}$ and $v_{\rm div}$ are the initial and final volume of a cell over the course of a cell cycle, then the single-cell elongation rate is defined as $\lambda = 1/\tau \ln (v_{\rm div}/v_{\rm birth})$. By taking the average of $\lambda$ over all cells in each experiment, we obtained the average single-cell elongation rates, $\langle \lambda \rangle$. 

 \begin{figure}[h!]
\includegraphics[scale=0.8]{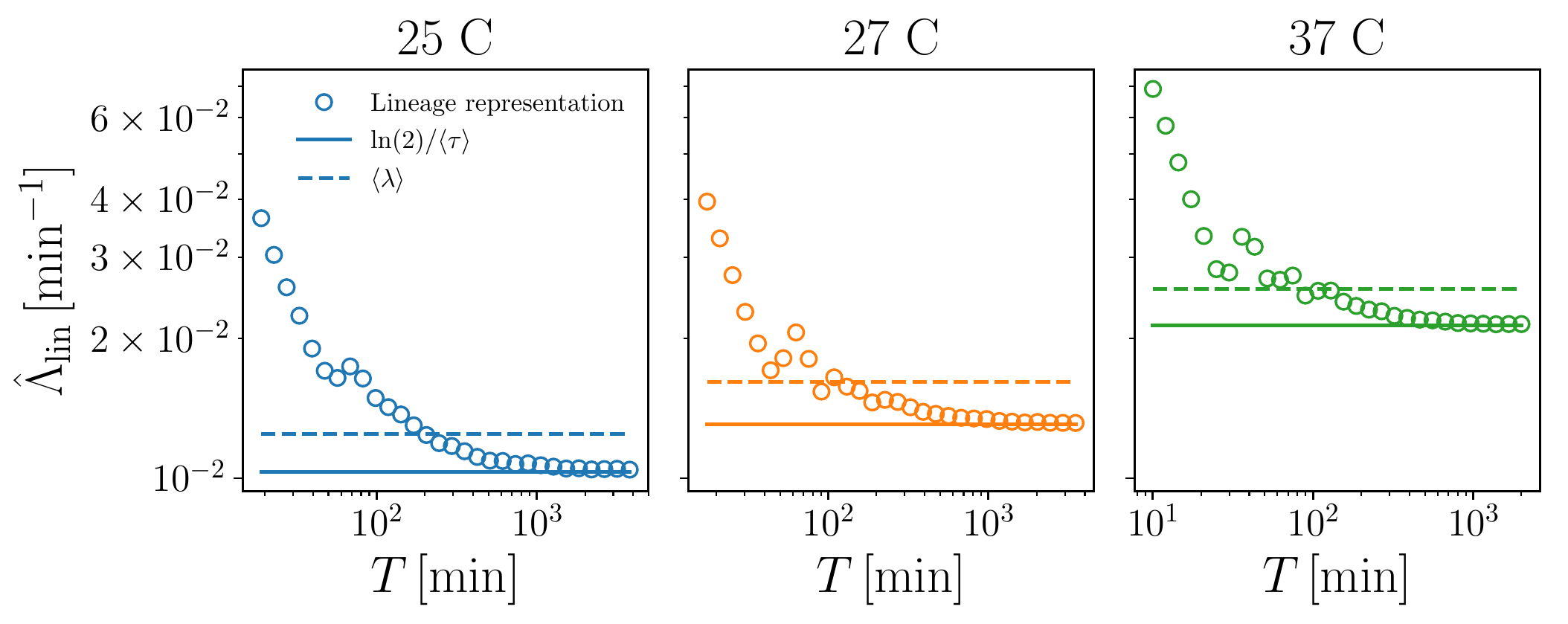}
\caption{The application of the lineage representation to experimental data. Each panel corresponds to an experiment done in a different temperature (indicated above the panel). Error bars for $\ln(2)/\langle \tau \rangle$ and $\langle \lambda \rangle$ can be computed from the standard error and are small enough to be invisible in these figures. 
}\label{fig:SI4}
\end{figure}

The results of our analysis are shown in Figure \ref{fig:SI4}. Here, we see that for large $T$,  $\hat{\Lambda}_{\rm lin} \approx \ln(2)/\langle \tau \rangle$; therefore, the lineage algorithm is not resolving the higher order effects on the population growth rate.  Comparing the trajectories of $\hat{\Lambda}_{\rm lin}$ with those from the insets in Figure \ref{fig:2} of the main text and Figure \ref{fig:SI3}, we see a qualitative  agreement. In particular, they seem to decrease initially before fluctuating around a constant, before decreasing to $ \ln(2)/\langle \tau \rangle$. This indicates that our theory indeed captures the performance of the lineage representation on real data, although without the corresponding population growth rate measurements, we cannot make any claims about the accuracy of the growth estimates.

\end{appendix}

\end{document}